\documentclass[10pt,preprint2]{aastex}

\newcommand{\et}{{\it et~al.\,}}

\begin{document}
\title{Phase-resolved Crab Studies with a Cryogenic TES Spectrophotometer}
\vspace{-0.4cm}

\author{Roger W. Romani\altaffilmark{1,2}, A.J. Miller\altaffilmark{3}, B. Cabrera}
\affil{Dept. of Physics, Stanford University, Stanford CA  94305-4060}
\and
\author{S.W. Nam and John M. Martinis }
\affil{National Institute of Standards and Technology, Boulder, CO 80303}

\altaffiltext{1}{also ATNF, CSIRO, Epping, NSW}
\altaffiltext{2}{Guest Observer, McDonald Observatory }
\altaffiltext{3}{Present Address NIST, Boulder, CO 80303}

\begin{abstract}
	
	We are developing time- and energy-resolved near-IR/optical/UV
photon detectors based on sharp superconducting-normal transition
edges in thin films. We 
report observations of the Crab pulsar made during prototype testing
at the McDonald 2.7m telescope with a fiber-coupled transition-edge sensor
(TES) system. These data show
substantial ($\delta \alpha \sim 0.3$), rapid variations in the spectral
index through the pulse profile, with a strong phase-varying IR break
across our energy band. These variations correlate with X-ray spectral
variations, but no single synchrotron population can account for the full 
Spectral Energy Distribution (SED). We also describe test
spectrophotopolarimetry observations probing the energy dependence of the 
polarization sweep; this may provide a new key to understanding 
the radiating particle population.

\end{abstract}


\section{Introduction }

	Transition-Edge Sensor (TES) detectors are showing great promise as
fast bolometer arrays in astronomical applications from the sub-millimeter
%({\it SCUBA device reference}) 
through the X-rays (Wollman, \et 2000).
In the near-IR/optical/UV range these devices offer good broad-band 
quantum efficiency (QE),
high time resolution and modest energy resolution and saturation count-rate
(Cabrera \et 1998, Romani \et 1999). Along with competing cryogenic
technologies (e.g. Superconducting Tunnel Junction devices, STJs;
Perryman, Foden \& Peacock 1993, Perryman \et 1999) these sensor arrays
offer the potential for important new capabilities, particularly for the
study of rapidly varying compact object systems. We are developing sensitive
TES spectrophotometer systems for such applications and report here on
test observations of the Crab pulsar with a prototype fiber-coupled array.

	The basic principles and present performance of thin-film 
tungsten (W) TES devices have been recently summarized in 
Miller \et (2000).
In brief, the systems routinely achieve energy resolutions of $\sim 0.15$eV
at $\sim 3$eV, photon arrival time resolution of $\sim 300$ns and single
pixel count rates of $\sim 30$kHz without serious pile-up problems. 
{\it En route} to
a high sensitivity, general purpose camera suitable for faint object
high speed spectrophotometry, we have scheduled a number of astronomical
demonstrations. Observations at the McDonald Observatory 2.7m Harlan J. Smith
telescope were made in February 2000 with a system that incorporated a 
number of substantial improvements over the single pixel, fiber-coupled
system used for the first astronomical observations described in Romani, 
\et 1999. 

\section{Experimental Apparatus}

	The detector system used in these observations employs a 
$6 \times 6 $ array of W TES pixels on a Si substrate with a $23\mu$m pitch, 
fabricated by our group using the Stanford Nanofabrication Facility.
For ease of fabrication $1\mu$m wide Al leads, the `voltage-bias rails',
were connected in the pixel plane which means that the
fill factor decreases towards the sides of the array as the active tungsten 
area is reduced to accommodate these leads (Figure 1). The rails,
superconducting during operation, are covered by an extension of the active
W thin film of the TES
pixel, providing electrical connectivity. Rails and pixels are
separated by $1\mu$m gaps. Because of the underlying Al, photons absorbed
above these rails in fact couple a larger fraction of the the deposited energy
to the W $e^-$ system
than those absorbed directly on the adjacent TES. These `rail hits' thus
increase the array fill fraction, but produce a `satellite peak' in the 
energy PSF, which complicates the spectral analysis (\S 3).  The data 
system used for these tests had six read channels 
so in practice a $2 \times 3$ sub-array along the 
midline was used. These uniform
$20 \mu$m$\times 20\mu$m pixels gave an effective fill factor of 96\%. 
The geometrical peculiarities of the arrays used in these measurements 
are not intrinsic to the TES system.
New arrays have masks to eliminate the rail events. Buried wiring and
focusing collimators have been designed that can further maintain
uniform pixel size and high effective fill factors over substantially larger
TES arrays.

\begin{figure}[!h]
%\epsscale{.5}
%\plotone{f1.eps}
\includegraphics[]{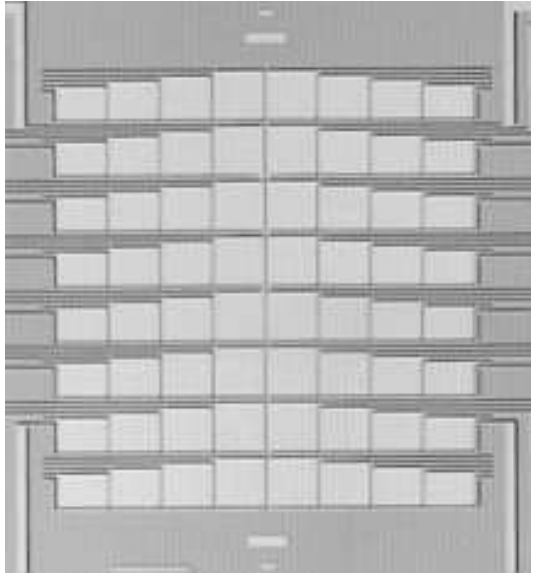}
\caption{TES Array image. For these observations, we recorded data
from the 4 pixels along the midline best illuminated by the focussed fiber.
The outermost pixels of the $8\times 8$ configuration are not wired and
serve only to mask the substrate.}
\label{array}
\end{figure}
	
	In operation, the Si substrate is cooled well below the 
$\sim 100$mK W transition
temperature, while the TES sensor is biased in a circuit with
a fixed voltage. The resulting current allows the W electron system to
self-heat into the middle of the superconducting-normal transition.
The TES sensors are thus operating in 
a regime of strong negative electrothermal feedback (Irwin 1995).
A major advance of the present apparatus is the first 
implementation of a NIST-developed digital feedback system which
simultaneously linearizes the SQUID ammeters and functions as the data 
acquisition system.  The digitized feedback signal monitors
the TES device current. The data stream is processed through an FPGA 
with adjustable peak shaping and baseline-restore algorithms. In addition
to accurate GPS-stamped arrival times and pulse height measurements for 
each photon, the system provides a number of quality control indicators
including adjustable pile-up flags. Four of the pixels were read with
independent digital channels of this new design; these provide the
photon set analyzed in this paper.
This system is directly scalable to large numbers of read channels
and can be modified to allow multiplex reading of several pixels per read
channel (Chervenak \et 1999). 

	There are two serious challenges in coupling these TES detectors
to a typical ground-based
astronomical telescope. One is to filter the incident beam
so that the large flux of thermal photons (at  $\lambda > 2\mu$m)
from the warm optics does not saturate the TES sensor system, which has a
modest maximum count rate.
The second challenge is to maximally couple the beam of
a large telescope to our present small (effectively 46$\mu$m) detector array.
In the system described here
we used a focal reducing train of fiber and imaging
optics (Figure 2). Given the relatively large plate scale
($232 \mu$m/arcsec at the f/17.7 Cassegrain focus of the 2.7m telescope) 
and best expected image of 1-1.5$^{\prime\prime}$ FWHM we adopted a 400$\mu$m
(1.7 arcsec) entrance aperture and fed the light through a 2:1 reducing 
taper. This taper, along with the $\sim 25$m run of 200$\mu$m-core fiber
to the refrigerator bulkhead, were fabricated from low OH, low impurity
(Polymicro FSU) Si/Si step-index 
fibers, providing both good IR and UV transmission. 
At the bulkhead of the cryostat, a portable $^3$He/$^4$He
dilution refrigerator, an ST-connector vacuum feedthrough
passed the light to a $\sim 5$m length of high OH `wet' 200$\mu$m core fiber.
The bulk of this fiber was spooled in thermal contact with the 1K stage
of the cryostat. In this way the strong OH absorption bands of the 
wet fiber (Humbach 1996) 
acted to provide a cold filter with effective blocking beyond
1.7$\mu$m and several strong shorter wavelength absorption bands
(e.g. at 0.95$\mu$m). These absorptions correspond to wavelengths
of strong atmospheric OH emission. The system thus provided 
transmission in the astronomically interesting J and H bands as well
as reasonable throughput to 400nm where scattering in the fibers
began to compromise transmission.

\begin{figure}[!ht]
%\epsscale{.67}
%\plotone{f2.eps}
\includegraphics[scale=0.35]{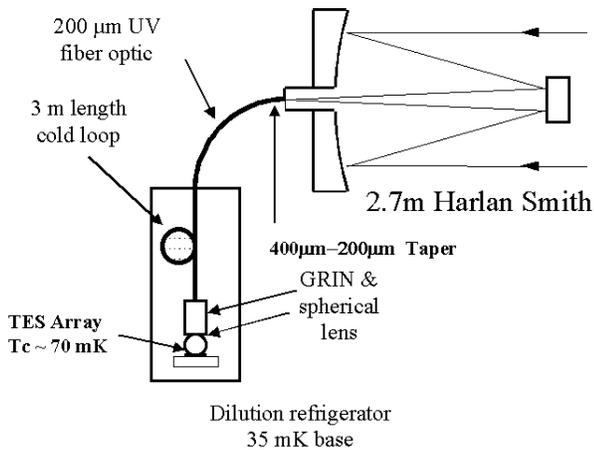}
\caption{Schematic of TES Spectrophotometer system}
\label{Schem}
\end{figure}

	To illuminate the sensor array at the cold stage, we collimated
the fiber beam with a 1.8mm 1/4 pitch gradient index (GRIN) lens and then
focused with a 1.0mm diameter spherical lens to provide the best coupling 
to the central $2\times 2$-pixel section of the TES array. 
Despite inevitable focal
ratio degradation in the fiber train, the final $\sim f/1$ focus with
modest chromatic aberration ensured that we had reasonable coupling into
our central 160$\mu$m$^2$ of active W TES. We measured the absolute
quantum efficiency of the system from the cryostat bulkhead to
final recorded photons, including the $\sim 50\%$ absorptivity of the 
grey tungsten surface, using a calibrated 850nm light source,
calibrated attenuators and an absolute power meter. These measurements
determined a total 850nm QE of 0.097. This was confirmed by comparison
with CCD integrations, using cameras of known QE. The `rail' hit photons
contribute an additional 0.01 to the system QE.  Inevitably, the 
fiber train to the telescope incurs additional losses. Unfortunately, these 
were exacerbated by a defect in the ST couplers of the taper section.
As a consequence, the fiber section imposed a factor of $\approx 2.5$ decrease 
in the peak throughput, relative to the back end of the telescope.

\section{Calibration}

	The energy scale for each pixel was set by interrupting the observations
every few hours to take short $\sim 60$s exposures of a calibration lamp 
passed through a grating monochrometer and fiber fed to the TES array. 
Typically we integrated with the 1$^{st}$ order
set at $\sim$0.5eV, 1.0eV and 2.0eV. Observations of the order peaks provided
a quick calibration of the system energy scale and non-linearity.
In addition, to monitor any slow drifts in the energy gain as might occur
due to baseline temperature drifts in the dilution refrigerator, a series
of calibrated `heat pulses' with a range of energies were introduced to
the W $e^-$ system on a regular cadence throughout the data exposures.
These pulses produce photon-like signals of known arrival time and amplitude
interleaved through the data stream. These heat pulses were excised
and monitored for any sign of gain drift. In practice, 
operation was sufficiently
stable that no detectable drifts were generally seen in the heat pulse
monitor.  The monochrometer calibration spectra were used to form the 
response matrices $R_k(E_i,\epsilon_j)$ across detector energy bins $\epsilon_j$
to the incoming photon energies $E_i$ for each pixel $k$.

\begin{figure}[!h]
\includegraphics[scale=0.35]{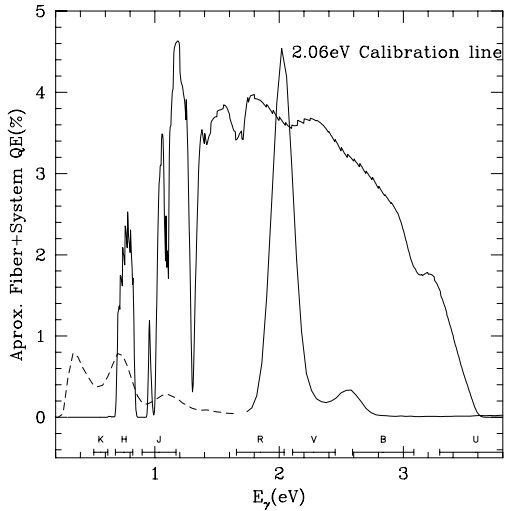}
%\epsscale{.7}
%\plotone{f3.eps}
\caption{The estimated total quantum efficiency of our system (fiber+detector)
at the back end of the telescope. Atmospheric absorption, telescope
inefficiencies and aperture losses give a typical peak effective
efficiency (relative to flux at the top of the atmosphere) of 
$\sim 1$\%.  For comparison
the count spectrum observed from a 2.06eV setting of a grating monochrometer
is shown. The low energy events (dashed line) include a rising continuum
due to substrate (Si) hits,
as well as a thermal IR H-band peak from the warm optics and lamp.}
\label{QE}
\end{figure}

	To complete the spectrophotometric calibration, we observed the
Bruzual-Persson-Gunn-Stryker spectrophotometric standard BD+30 2344
at low airmass during each observing night of the run. Although
convenient for its well calibrated absolute IR-optical-UV
spectral energy distributions, at $m_V = 10.2$ the star had to be observed with
substantial de-focus to avoid detector non-linearities at high count rates.
We therefore also observed the $m_V = 14.9$ blue KPNO flux standard PG 0939+262
(Massey \et 1988) to allow an improved estimate
of the system UV response and a check on the point-source total efficiency,
including aperture losses. The fiber filter has strong absorption bands
whose width is comparable to or smaller than our energy resolution. Also
for steep red spectra, the satellite `rail hit' peak at 1.26 times the main
peak energy contributes significantly to the count rate. This made it 
impossible to extract the flux scale by directly dividing the predicted 
standard star counts by the observed channel counts.  Instead
the calibration star spectrum was convolved with the response function, the 
fiber absorption
curve measured at high resolution, and the W absorption spectrum to
produce model calibrator count spectra. Comparing in the data space, an
effective area function varying slowly with energy
was fit to model the additional
focus and absorption losses compared to stellar fluxes at the top of the 
atmosphere. Combining these factors, we derive 
model efficiency functions $E_k (\epsilon_j)$ 
for each active pixel $k$. After summing these we have a total 
(detector+telescope+atmosphere) efficiency for the system; this peaks at
$\sim 1.2$\%. In Figure 3 we correct this efficiency for the estimated
% final normalization from the Crab spectrum (assume this is best track)
atmospheric extinction, telescope inefficiencies and aperture losses
to produce the effective response of the system (fiber optics + detector) at
the back end of the telescope.  For comparison, the energy ranges corresponding 
to Johnson broad-band colors are indicated. Variation between the calibration 
sources indicate that the systematic errors in the flux scale are $< 10\%$ 
for the range 1-2.5eV. In the blue, the calibration spectra have lower S/N and
larger contribution from the rail hit events, so the systematic errors
in the flux scale rise, probably reaching 30\% above 3.5eV. Comparison
with archival Crab data below support the idea that our flux scale is
systematically high by approximately this amount in the U band.
Imprecision in the estimates of the telescope and aperture losses also 
produce uncertainties exceeding 10\% in the absolute flux scale.

\section{Crab Pulsar Spectrophotometry}

	We observed a number of compact object sources, both accretion
powered binaries and isolated spin-powered pulsars. The relatively bright
($m_v \sim 16.6$) Crab pulsar was used to tune and characterize the system 
performance. 

	Our GPS-stamped Crab photon arrival times were reduced to the
solar system barycenter employing the widely used
TEMPO pulsar timing package (from pulsar.princeton.edu/tempo). We used
the monthly Crab pulsar radio ephemeris published by Jodrell Bank
(Lyne, Pritchard \& Roberts 1999) to phase the photon arrival times.
The light curves from our four nights of observation are phased within
$\sim 10\mu$s; the absolute phase agrees with the predicted infinite
frequency pulse arrival time (referenced to the peak of the main pulse)
to within $\sim 30\mu$s. This absolute precision is rather better than the
nominal $\sim 100\mu$s accuracy of the published monthly ephemeris
and is completely
adequate for our purposes, allowing a detailed comparison of our
phase bin spectra with other (e.g. space-based) data sets. We focus
here on a relatively long (3500s) coherent data set taken on February 6, 2000
(MJD 51581)
for which the system parameters were stable and the observing efficiency was
high. This data set provides much of our total S/N for
the Crab pulsar and represents a unique time- and energy-resolved measurement
of this well-studied object. A sample optical (1.5eV-3.5eV) light
curve is shown for reference in Figure 4.

\begin{figure}[!h]
\includegraphics[scale=0.35]{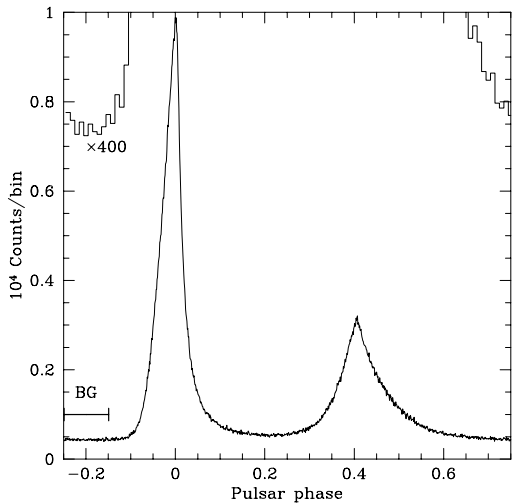}
%\epsscale{.7}
%\plotone{f4.eps}
\caption{A sample 1.5eV-3.5eV (3550 - 8250 \AA) 
$10^3$-bin light curve from 3500s of
TES Crab data. Phase 0 is referenced to the main peak of the radio ephemeris.
The upper histogram shows the same data at 100 bins (zero suppressed),
magnified to show the baseline variation. Our selected background
region is shown (label BG).
}
\label{Lightcurve}
\end{figure}

	There is appreciable structure in our light curve baseline even 
at pulse minimum. Previous studies also infer magnetospheric emission
throughout the pulse period (e.g. Sanwal \et 1998). Recently
Golden, Shearer \& Beskin (2000) have argued from MAMA imaging data that 
the phase interval 0.75-0.825 represents the DC pulse minimum 
with $\sim 0.4$\% of the integrated pulse emission. Our data show significant 
structure in this region, but given the substantial background in our
our 1.7$^{\prime\prime}$ aperture, we select the phase interval 
$\phi_{BG}=0.76-0.87$ for our `background' spectrum, recognizing
that there will be a small $< 1$\% magnetospheric contamination; this
will be dominated by Poisson fluctuations in nearly all of our extracted 
spectra.

\begin{figure}[!h]
\includegraphics[scale=0.54]{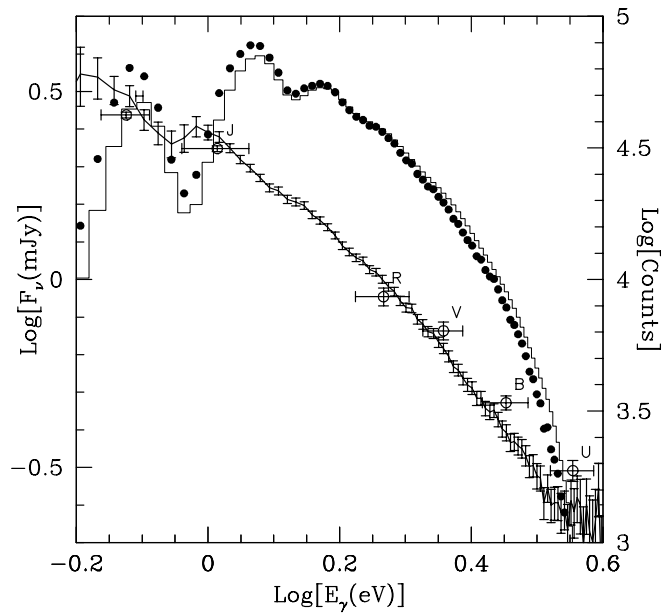}
%\epsscale{.7}
%\plotone{f5.eps}
\caption{Phase-averaged TES Crab spectrum (full line with error flags)
compared with broad-band colors. Predicted (extincted $\alpha_\nu = 0.11$;
line histogram) and observed (filled dots) count spectra are shown
for comparison (Right scale).
}
\label{Phaseav}
\end{figure}

	In Figure 5 we show the phase-averaged spectrum after background
subtraction, convolution through the response matrix and application 
of the flux calibration solution of \S 3. To maximize the independence of
the spectral bins, we first compute the expected count spectrum of the best-fit 
simple absorbed power-law model and then derive the Crab fluxes by 
multiplying the model flux by the ratio of the observed to predicted
counts for each channel. This `pre-whitening' avoids spreading 
fluctuations in the observed count bins between final spectral bins
and ensures that sensitivity to errors in the response matrices are
of second order. Nonetheless, the derived spectrum represents only
one model (the smoothest) that fits the count data. The error bars
shown are Poisson only, derived from the independent counts in the energy bins.
For comparison, we plot optical/IR phase-averaged 
broad-band photometry tabulated in Eikenberry \et (1997). We also show
the phase-averaged spectrum fit by Sollerman \et (2000) to optical/UV
data as the solid histogram -- this is an $\alpha_\nu=0.11$ powerlaw
subject to E(B-V)=0.52, R=3.1 interstellar reddening -- displayed as a 
predicted count histogram. This may be compared with the observed spectrum
(solid dots). Notice that, in agreement with the IR photometry of Eikenberry 
\et (1997), this powerlaw under-predicts the J/H flux. 
	
	It is evident that our derived spectrum shows a somewhat steeper 
(redder) optical spectrum than the Sollerman \et power-law fit and also lies
below the B \& U colors of Percival \et (1993).  This supports the suspicion 
of a systematic over-estimate of our sensitivity function
above 2.5eV (possibly due to incomplete modeling of the rail-hit
component of the PSF and/or pile-up). The notch in the spectrum at 
$E_\gamma \sim 0.9$eV is also likely an artefact caused by the rapid 
variation in the system sensitivity at the OH absorption band.
Interestingly, we find that our $E\ge 1$eV spectrum is however very well 
modeled by an extincted $\alpha=-0.2$ power law,
which is actually the same spectrum found originally by Oke (1969). The
true spectral index is evidently quite sensitive to the reddening parameters
and energy range chosen and, indeed, the optical/UV spectrum of
Sollerman \et shows appreciable curvature. However, given the
mismatch with the broad-band photometry in the blue, slight flux
calibration systematics are the likely culprit.

	We can, of course, study spectral variations in the Crab by
assuming that the phase-averaged spectrum is smooth and using our observed
total pulsar spectrum as the flux calibrator. This differential
measurement allows a much
improved removal of small scale variations in the sensitivity function
and provides robust detection of any spectral phase variations. Figure
6 shows a selection of phase bin spectra, referenced to the total
(phase-averaged) spectrum. There are evidently subtle but
continuous variations in the broad-band spectral index and a distinct
change in the behavior in the near-IR band (see, for example the break
at $\ga 1$eV in the leading edge of the main pulse, spectrum 1). The spectra
show little fine structure except at the strong fiber absorption
bands. Smooth continuum variations dominate, as expected for a 
broad-band emission process such as optical synchrotron. Errors are computed
by propagating Poisson fluctuations in both the background
subtracted spectra and the reference (phase-averaged) spectrum.

\begin{figure}[!h]
\includegraphics[scale=0.65]{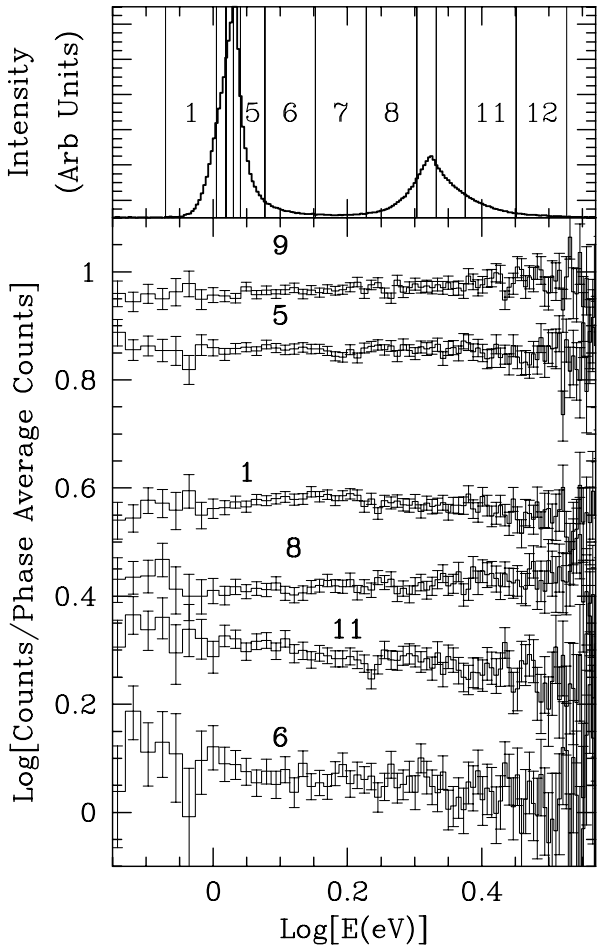}
%\epsscale{.52}
%\plotone{f6.eps}
\caption{Phase bin count spectra. The log of the photon flux is plotted
against log($E_\gamma$) for several phase bins from the 13 contiguous
bins indicated in the panel to the right.
}
\label{Phasesamp}
\end{figure}

	We can measure these smooth variations in smaller $\phi$ bins
by fitting. Figure 7 shows a double power-law fit with a fixed
break at 1.3eV. To obtain these spectral indices, we correct by
a mean (phase-averaged) powerlaw. In view of the calibration uncertainties
above, we have chosen to reference the spectra using a phase-averaged
index of $\alpha=-0.1$. The spectral index fit errors are derived from
the Poisson spectral bin errors above.

\begin{figure}[!h]
\includegraphics[scale=0.45]{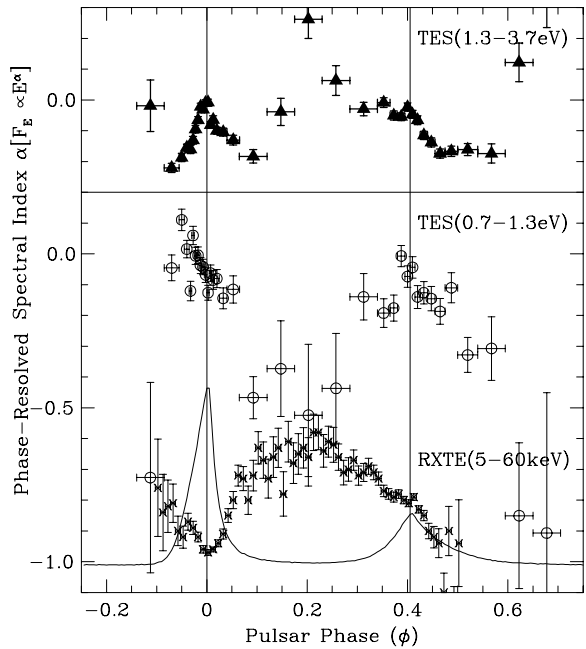}
%\epsscale{.7}
%\plotone{f7.eps}
\caption{Spectral index phase variations. Fits to optical (triangles, 
Upper panel) and near-IR (open circles, Lower panel) bands are shown. 
Note the rapid, anti-correlated
variation in the main peak, indicating a varying spectral break.
For comparison, an optical light curve and X-ray {\it RXTE} 
(Pravdo, Angelini \& Harding 1997) phase-resolved spectral indices are shown
in the lower panel.
}
\label{Phasevar}
\end{figure}

	The data show a rapid sweep in the spectral index, especially
in the main peak. An optical light curve and peak phase marks are
provided for reference.
At the leading edge of the  main peak the IR index
does not reverse, but the spectrum becomes increasingly blue. This
indicates strong break or depression of the IR at the leading edge of the
main pulse. This is the origin of the IR variation in the leading pulse
half-width noted by Eikenberry, \et (1997).
While the phase bins and error flags are large in the 
bridge between the pulses, the optical and IR indices also differ 
strongly in this region, with a steep red IR spectrum passing  to
a much harder component in the blue. This increase of the 
bridge flux at high energies is in accord with the strengthening of
the bridge emission in the X-ray and $\gamma$-ray bands.

	Phase-dependent spectral indices were first noticed in the Crab
in the X-ray band (Pravdo \& Serlemitsos 1981). The reference {\it 
Rossi X-ray Timing Explorer (RXTE)}
data above and more extensive 
BeppoSAX opservations reported by Massaro \et (2001)
show rapid variation in the hard X-ray power-law spectrum, especially
near the first peak. The phase coincidence of the optical/IR variations
imply a common origin, but there is evidently no simple relation between the
indices in the two band. At higher $\gamma$-ray energies limited count rates
do not allow the fine phase bins used in the X-ray analysis, but a correlated
spectral variation is seen across the Crab pulse profile (Fierro 1995).

	In the optical, many authors starting with Muncaster \& Cocke (1972)
have used used time-resolved color photometry to suggest that the bridge
region is bluer than the pulses. This is seen most clearly in the
largely unpublished UBVR observations reported in Sanwal \et (1998).
Recently, Golden \et (2000) have reported a {\it reddening}
in the bridge region, based on UBV photometry, but this does not accord
with other recent measurements. Note that with the exception of the Sanwal
\et (1998) data, limited count statistics forced all observations to
use very coarse phase bins, typically summing over the entire pulse  
or interpulse. The $\alpha$ sweeps we see are located largely within
these components; this is a likely explanation for the non-detection of
spectral variation in several studies
(e.g. Carraminana, Cadez \& Zwitter 2000, Perryman \et 1999).
Thus the optical $\alpha$ variations through the pulses are seen here clearly
for the first time.

	The simultaneous IR measurements of this data set also provide
a new window on the pulsar physics, given the rapid
variation in the spectral break below $\sim 1$eV. Simultaneous optical/IR
spectra were first reported in Romani \et (1999). Recent 
simultaneous dual band (V \& H) light curves have also been reported by 
Moon \et (2001), but the low efficiency and limited S/N of these test data
likely preclude measurement of the subtle spectral variations.
Although some interesting indications
of spectral variations can clearly be seen in the broad-band (non-simultaneous)
IR photometry of Eikenberry \et (1997), the data reported above are the first
to show these fine-resolution spectral variations extending into the infrared.
While the variations are highly significant, longer integrations with better
S/N are certainly desirable. Nonetheless, these data already provide
some constraints on the radiation processes, which we describe
briefly in \S 7.

\section{Crab Photon Statistics}

%perhaps mention claimed 60s precession period -- Albetro \& Co (2001)

	At our roughly kilohertz count rates, we detected $\sim$25
photons in each main pulse, giving us substantial sensitivity to
photon correlations. In the radio most pulsars exhibit a Gaussian
or log-normal
distribution of pulse intensities. The  Crab, however, also produces 
occasional radio pulses with up to $\sim 10^3 \times$ the mean pulse energy. 
These 'Giant pulses' are short duration bursts of flux, appearing
randomly in narrow windows of pulsar phase.  Lundgren \et (1995)
have searched for and placed limits on any correlation between giant 
radio pulses and $\gamma$-ray emission, while Patt \et (1999) have found
that no giant pulses are present in the X-ray band. Percival \et (1993)
have also found no non-Poisson pulse-to-pulse variations in the optical flux
of the individual pulse components.  Lacking simultaneous 
radio data, we examined our data for any short-time scale photon correlations.

	We searched for non-Poisson photon distributions by predicting the
expected counts in a range of lag bins following each observed photon.
Because each detector has a small $\sim 10\mu$s dead time associated with
the recovery from a photon pulse, we searched the {\it cross}-correlations
between our pixels, allowing us to extend the analysis to microsecond times.
To test the null Poisson hypothesis, we need
an accurate estimate of the expected count rate in the face of the varying
pulsed emission and background in our aperture. This was obtained by
forming a 100-bin light curve  for each second of data and fitting the
counts to a background and pulsed flux amplitude. For each photon arriving
during this second, we used the two fit amplitudes to {\it predict} the
photon arrival rate in the other three detectors on a range of timescales.
The model flux in each lag bin was determined by an integration through
the appropriate phases of a high-resolution ($10^3$bin Figure 4) light 
curve from all the data in a selected energy range, scaled to
the instantaneous detected flux. The observed counts can be compared to
the Poisson expectation for photons of a given energy range on various 
timescales and over various windows of pulsar phase. No strong energy
dependency was noted, so we show the statistics of the full low background
1-4eV energy range.

\begin{figure}[!h]
\includegraphics[scale=0.52]{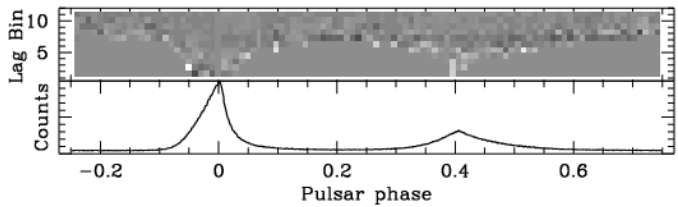}
%\epsscale{1.}
%\plotone{f8.eps}
\caption{Ratio of average rate to Poisson rate for $E_\gamma \ge 1$eV. 
The grey scale image shows the
ratio in independent lag bins, spaced as even octaves (vertically), vs. 
pulse phase (horizontally). Light shows high correlation and dark shows
sub-Poisson counts. Bins with $< 25$ counts are set to 1.
}
\label{Pois}
\end{figure}

	Figure 8 shows as a gray intensity scale the observed count rate
in units of the Poisson predicted rate in 100 bins of pulsar phase.
The vertical axis of the grey scale shows correlation bins covering time
lags $\delta \tau = t_{samp} (2^N - 2^{N-1})$, where $t_{samp}= 0.31\mu$s.
$N$ runs from 1 to 11 ($\tau=0.3\mu$s to $635\mu$s) in the
gray scale plot shown. The lag
bins are thus completely statistically independent. A significant
correlation (or anti-correlation) on short time scales would be expected 
to show as a coherent bright (or dark) region at the appropriate phase,
fading to grey in longer timescales. The data show fluctuations in the
individual amplitudes when the counts are low. There is a $\sim 10-15$\%
enhancement on timescales $\tau \sim 5-10\mu$s leading the first peak by 
$\sim 1$ms; this is of $\sim 3\sigma$ significance. There is also
a $\sim 30$\% enhancement at $\tau \le 3\mu$s leading the second peak,
but this is only $\sim 2 \sigma$ in the most significant lag bin.
A few percent
decrement at lag $N=7$ appears correlated with noise in the 33$\mu$s 
bins in the fine template. All other significant features are smaller
than 1\%. We conclude that our photon arrival times are consistent with
Poisson and that there is no evidence for strong correlations on short 
timescales, such as giant pulses, at optical energies.

\section{Crab Spectrophotopolarimetry}

	Ultimately, complete exploitation of astronomical fluxes requires
time-resolved spectrophotopolarimetry. The strong, rapidly varying
polarization of the Crab pulsar
makes it an attractive test target. Phase resolved
polarimetry has been presented in the optical (Smith \et 1988) and the
near UV (Smith \et 1996), which give a basis for comparison. The double
position-angle sweep reported in these papers has been shown to
be a good quantitative match to a relativistic outer magnetosphere 
version of the rotating vector model
by Romani and Yadigaroglu (1995). In this model emission from a range of 
altitudes forms caustics in photon arrival phase that define the pulses.
Further, a combination of two distinct emitting regions can contribute to the 
interpulse phases. If the radiating populations (and hence spectral
indices and breaks) differ between these regions we would expect energy
dependence of the polarization fraction.

	As a test, we observed the Crab pulsar on February 7, 2000 (MJD 51582)
with a simple polaroid filter placed before the focal 
plane. We integrated at two full cycles of position angle with the following
exposures: $0^\circ$ (1370s,630s), $45^\circ$ (520s,690s), $90^\circ$
(740s,620s), $135^\circ$ (960s,680s) with angles referenced to the equatorial
coordinate system. Of course, varying background and
varying aperture losses between the exposures mean that the light curves must
be renormalized before forming $q=(I_0-I_{90})/(I_0+I_{90})$ and
$u=(I_{45}-I_{135})/(I_{45}+I_{135})$. To subtract the substantially
polarized nebular emission in our aperture, we employed our usual background 
phase window (0.76-0.87), following the prescription in (Smith \et 1996).
While residual pulsar emission at these phases
may be highly polarized (Smith \et 1988), the amplitude is small enough
that its contamination of the peaks and bridge region should be smaller 
than our statistical errors. An examination of the 
Smith \et (1988) data (kindly supplied in digital form by F.G. Smith)
showed a polarization minimum during the bright first pulse in the region
$\phi =$0.005-0.025, consistent with 0, within the error bars. This provides
a convenient exposure monitor based on the pulsed emission itself. These data
were taken in a broad optical band with the `RGO people's photometer'; we
synthesized this band in our data, selecting the appropriate energy bins 
and using the polarization minimum flux
as a monitor of the total Crab exposure at each polarization position angle.

\begin{figure}[!h]
\includegraphics[scale=0.42]{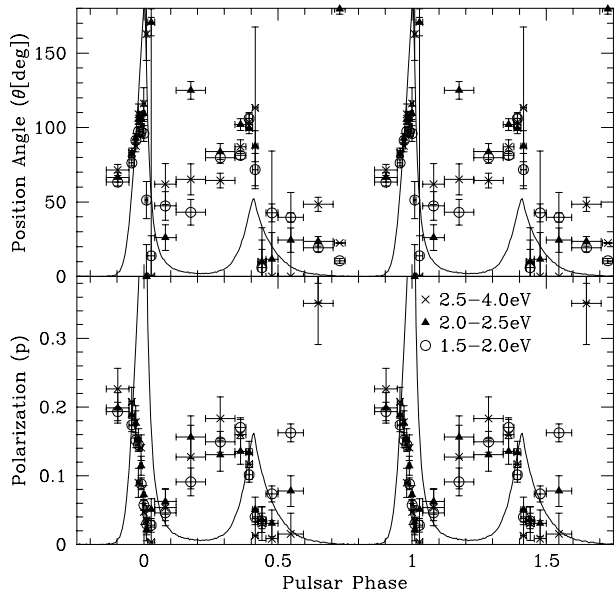}
%\epsscale{.7}
%\plotone{f9.eps}
\caption{Test Crab Spectropolarimetry in three optical bands. Two
pulse periods are shown for clarity. 
Above: Position angle. An optical light curve is included for phase reference.
Below: Linear polarization fraction.
}
\label{Polar}
\end{figure}

	In Figure 9, we show the resulting polarization fraction
$p=(q^2+u^2)^{1/2}$ corrected for the wavelength dependent polarization
efficiency of the filter (measured using monochrometer light passed through
crossed filters). We also corrected for the bias due to non-normal statistics
at small polarization, with the observed polarization $p_0$ corrected to
$p=p_0 [1-(\sigma /p_0)^2]^{1/2}$.  In the upper panel we show the
polarization position angle $\theta= 1/2 \times {\rm Arctan}(u/q)$.
The error flags shown are the propagated Poisson errors. Differences between the
two sets of position angle data were within the statistical errors, so these
were summed in the polarization amplitude and sweep shown. Limited statistics
restricted us to 18 phase bins through the pulse, so fine structure is
likely not fully resolved.

	The overall polarization behavior follows that measured in
previous optical/UV experiments quite closely. We show the optical data
divided into three energy bands. The rapid decrease in polarization and strong
position angle sweep of the first pulse appear quite energy independent.
There are small differences in the polarization and position angle of the
bridge and second pulse, but these are of marginal significance. Given that
the phase variations in the spectral index and break are strongest below 1.5eV
(\S4) we might expect the strongest spectral effects in the near-IR. Indeed,
our near IR data do appear to show significantly different polarization 
properties. Unfortunately, the increased IR background and difficulty
in calibrating the rapidly varying IR polarization response of the filter
preclude a quantitative comparison with the optical data in the present
experiment. Serious phase-resolved spectropolarimetry will
require imaging solutions for local background subtraction; as part of
our next-generation imaging system (see below), we plan to install
Wollaston prisms for imaging spectrophotopolarimetry over reduced fields.

\section{Spectral Variations and Conclusions}

	The phase variations through the Crab profile described above
clearly provide new tests of the physics of the optical/IR radiation.
The polarized IR through hard X-ray non-thermal emission from the Crab is 
generally held to be synchrotron radiation from a population of $e^\pm$
propagating and cooling in the outer magnetosphere (e.g. Lyne and Smith 
1990 and references therein; Romani 1996; Wang, Ruderman \& Halpern 1998).
The broad-band phase-averaged SED is smooth, with $\alpha$ varying from 
$-0.7$ 
(hard X-rays) through $-0.5$ (soft X-rays) to $\sim 0$ at optical frequencies.
There is evidently a peak in the near-IR and then the spectrum drops off 
dramatically at lower energies (Middleditch, Pennypacker \& Burns 1983),
although this important result definitely requires confirmation.

	Recently, Crusius-Waetzel \et (2001) have provided a
synchrotron interpretation that provides a plausible explanation of 
the soft-X-ray through IR
phase averaged SED in terms of a single $N \propto E_{e^\pm}^{-s}$ population
(presumably produced through an $e^\pm$ pair cascade in the outer magnetosphere)
with $s=2$. In the X-ray band, the emission is assumed to be classical
relativistic synchrotron radiation with $F_E \propto E^{-(s-1)/2}$;
this gives the observed $\alpha \approx -0.5$ spectrum observed for the 
powerlaw X-ray emission of the Crab and other young pulsars. However,
Crusius-Waetzel \et additionally argue that the $e^\pm$ population
should have a component with very low pitch angle $\Psi \ll 1/\gamma$,
for which the synchrotron emission formulae give $F_E \propto E^{2-s}$,
when averaged over the radiating beam.
For the Crab, they infer $\gamma \sim 10^2-10^3$ and $\Psi < 10^{-3}$,
and connect the $\alpha \approx 0$ spectrum with that seen in the optical. 
Finally,
they suggest that below $1$eV the particles dominating the emission
spectrum have even lower $\gamma$, so that the observed pulse width
$\Delta \phi \sim 0.01$ 
samples only a subset of the $1/\gamma$ radiation cone and argue that
this leads to a $F_E \propto E^{4-s}$ spectrum. This index is taken to be,
at least in part, an explanation of the steep rise in the mid-IR.

\begin{figure}[!h]
\includegraphics[scale=0.45]{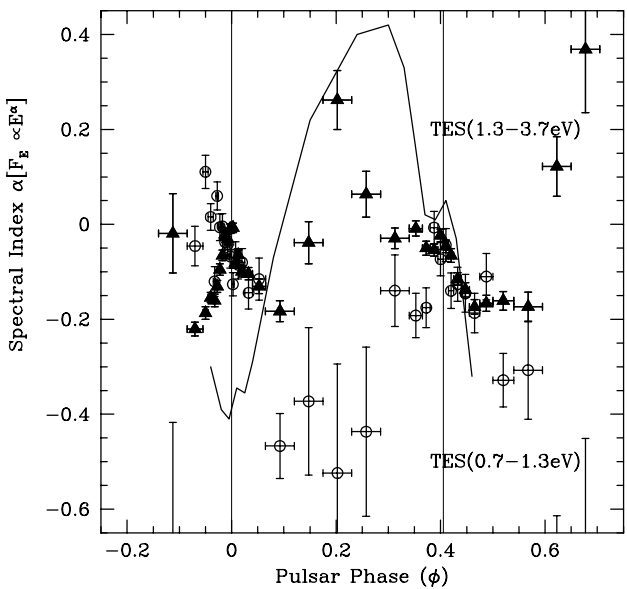}
%\epsscale{.7}
%\plotone{f10.eps}
\caption{Low pitch-angle prediction for the optical spectral index
variation (full line) compared with optical and IR TES data.
}
\label{Xpred}
\end{figure}

	If we take the X-ray spectral index to be a {\it local}
measurement of the electron powerlaw $s$ then this picture predicts
the correlated optical spectrum in the low pitch angle limit. In Figure
10, we show the TES data compared with this prediction, based on the
0.1-4.0\,keV data of Massaro \et (2001). Clearly this phase bin
application of the low pitch-angle picture is not adequate: while the
index and trend for the second pulse are reasonably well predicted,
the spectra of the bridge region and, especially, the first pulse show
no match to the extrapolations of soft X-ray data. When one considers
that the spectral index in all phase bins steepen into the hard X-rays,
then one must conclude that a simple application of powerlaw synchrotron
emission, even with a phase-varying $s$, is not adequate.

	One factor likely important in a more complete model is
the inclusion of the local beaming of the radiation cones along
the emission surface. Given the sharp pulses and the observation
at a single line-of-sight, the radiation must differ from the emission-cone
averaged synchrotron expressions above. Models including these
effects are sensitive to the detailed geometry. As an example,
consider the strong IR break at the leading edge of the main pulse.
In the outer magnetosphere caustic picture of the pulse formation,
the leading edge of the main pulse arises from emission propagating
along the radiating surface for the longest distance. Hence, one would
expect this to be the region first showing self-absorption suppression.
While the overall IR suppression of the main peak reported by Penny (1982)
suggestive of uniform self-absorption is not seen in our data, the observed
leading edge suppression and steepening may be a more restricted
instance of self-absorption. Polarization measurements extended into
the IR will be a good discriminant between absorption effects and
emission breaks, as a restricted sampling of the synchrotron emission
cone leads to polarization signatures varying significantly in 
accord with the spectral breaks.

	We are developing a focal plane camera based on TES detector
arrays, with the potential for time resolved imaging spectral
polarimetry. As the data of Golden \et (2000) and Perryman
\et (1999) show, the absolute photometry and background
subtraction of such true imaging measurements can provide much
improved isolation of faint pulsar phenomena. With the clean
monoenergetic energy PSF of a `masked' TES array, improved system QE
and improved energy resolution, observations of the Crab (and other
fainter pulsars) should allow a detailed examination of the phase-resolved
phenomena discovered in the exploratory measurements reported here.

\begin{acknowledgements}
This work was supported in part by grants from NASA (NAG5-3775 and NAG 5-3263),
from the DOE (DE-FG03-90ER40569), from the Stanford OTL fund and from 
the Research Corporation.
This work also made use of the Stanford Nanofabrication Users Network
funded by the National Science Foundation under award number ECS-9731294.
We thank Kent Irwin and
R. Welty of NIST for continued collaboration on TES and SQUID technology,
David Doss of McDonald Observatory for assistance with the observational
set-up and Colin Bischoff for assistance with the data analysis.
\end{acknowledgements}

%\vfill\eject


\begin{references}
\reference{} Cabrera, B., \et \, 1998, {\it Appl. Phys. Let.}, {\bf 73}, 735.
\reference{} Carraminana, A., Cadez, A. Zwitter, T. 2000, {\it ApJ},
{\bf 542}, 947.
\reference{} Chervenak, J.A., {\it et al.} 1999, {\it Appl. Phys. Lett.}, 
{\bf 74}, 4043.
\reference{} Crusius-W\"atzel, A.R, Kunzl, T. \& Lesch, H. 2001,
{\it ApJ}, {\bf 546}, 401
\reference{} Eikenberry, S.S., \et 1997, {\it ApJ}, {\bf 476}, 281.
\reference{} Fierro, J.M. 1995, PhD Thesis, Stanford University.
\reference{} Golden, A., Shearer, A. \& Beskin, G.M. 2000, {\it ApJ},
{\bf 535}, 373.
\reference{} Golden, A., \et 2000, {\it AA}, {\bf 363}, 617.
%\reference{} Gunn, J.E. \& Stryker 1983, L.L., {\it ApJ Suppl.}, {\bf 52}, 121.
\reference{} Humbach, O., \et 1996, {\it Journal of Non-Crystaline Solids}, 
{\bf 203}, 19.
\reference{} Irwin, K.D. 1995, {\it Appl. Phys. Lett.}, {\bf 66}, 1998.
\reference{} Lundgren, S.C. \et 1995 {\bf ApJ}, {\it 453}, 433.
\reference{} Lyne, A.G., Pritchard, R.S. and Roberts, M.E. 1999,
(http://www.jb.man.ac.uk/$\sim$pulsar/crab.html)
\reference{} Lyne, A.G. \& Smith, F.G. 1998, {\it Pulsar Astronomy}
(Cambridge:Cambridge).
\reference{} Massaro, E., Cusumano, G., Litterio, M. \& Mineo, T. 2001,
{\it AA}, {\bf },
\reference{} Massey, P., Strobel, K., Barnes, J.V \& Anderson, E. 1988, 
{\it ApJ}, {\bf 328}, 315.
\reference{} Middleditch, J., Pennypacker, C. \& Burns, M.S. 1983, {\it ApJ}, {\bf 273}, 261.
\reference{} Miller, A.J., \et 2000, {\it NIMPA}, {\bf 444}, 445.
%Cabrera, B., Romani, R.W., Figuer
\reference{}Moon, D.-S., Pirger, B.E. \& Eikenberry, S. 2001, PASP, 
{\bf 113}, 646
\reference{} Muncaster, G.W. \& Cocke, W.J. 1972, {\it ApJ}, 178, L13
\reference{} Oke, J.B. 1969, {\it ApJ}, {\bf  156}, L49
\reference {} Patt, B.L., Ulmer, M.P., Zhang, W., Cordes, J.M. \&
Arzoumanian, A. 1999, {\it ApJ}, {\bf 522}, 440.
\reference{} Penny, A.J. 1982, {\it MNRaS}, {\bf 198}, 773
\reference{} Percival, J.W. \et 1993, {\it ApJ}, {\bf 407}, 276
\reference{} Perryman, M.A.C., Foden, C.L., \& Peacock, A. 1993, {\it 
Nucl. Instr. Meth.}, A325, 319.
\reference{} Perryman, M.A.C., Favata, F., Peacock, A., Rando, N. \& Taylor, B.G.
1999, {\it AA}, 346, L30.
\reference{} Pravdo, S.H. \& Serlemitsos, P.J. 1981, {\it ApJ}, {\bf 246}, 484.
\reference{} Pravdo, S.H. Angellini, L. \& Harding, A.k. 1997, {\it ApJ}, 
{\bf 491}, 808.
\reference{} Romani, R.W. 1996, {\it ApJ}, {\bf 470}, 469.
\reference{} Romani, R.W., \et 1999, {\it ApJ}, {\bf 521}, L153.
\reference{} Romani, R.W. \& Yadigaroglu, I.-A. 1995, {\it ApJ}, {\bf 438}, 314
\reference{} Sanwal, D., Robinson, E.L. \& Stiening, R.F. 1998, {\it BAAS},
{\bf 30}, 1420.
\reference{} Smith, F.G., Jones, D.H.P., Dick, J.S.B. \& Pike, C.D. 1988,
{\it MNRaS}, {\bf 233}, 305.
\reference{} Smith, F.G., \et 1996, {\it MNRaS}, {\bf 282}, 1354
\reference{} Sollerman, J. \et 2000, {\it ApJ}, {\bf 537}, 861.
\reference{} Wang, F.Y.-H., Ruderman, M. \& Halpern, J.P. 1998,
{\it ApJ}, {498}, 373
\reference{} Wollman, D.A. \et 2000, {\it NIMPA}, {\bf 444}, 145.
\reference{}

\bigskip

%\reference{} Harris, L. 1956, {\it J. of Op. Soc. Am.}, {\bf 46}, 160.
%\reference{} Humbach, O., \et 1996, {\it Journal of Non-Crystaline Solids}, 
%{\bf 203}, 19.
%\reference{} Jakobsen, P. 1999, in Ultraviolet-Optical Space Astronomy 
%Beyond HST, eds J. A.  Morse, J. M. Shull, \& A. L. Kinney, ASP Conf. Ser., in press
%\reference{} Palik, E.D. 1985, ``Handbook of Optical Constants of Solids''
%(Academic Press:London).
%\reference{} Peacock, A. \et 1996, {\it Nature}, 381, 135
%\reference{} Peacock, A. \et 1997, {\it A\&A Suppl.}, 123, 581.
%\reference{} Romani, R.W., \et 1998, {\it BAAS}, {\bf 30}, 1266.
%\reference{} Welty, R. \& Martinis, J. 1993, IEEE Trans., Appl. Supercond. {\bf 3}, 2605.
%Clarke, R.M., Colling, P., Miller, A.J., Nam, S. \& Romani, R.W., 
% K.D. Irwin, Nam, S., Cabrera, B., Chugg, B. \& Young, B., {\it Rev. Sci Instrum.} {\bf 66}, 5322 (1995).
% Strecker, D.W. \et, Ap. J. Suppl. {\bf 41}, 501 
% (http://ra.stsci.edu/documents/SyG/SG\_50.html)
%Fazio, G.G., Ransom, S.M., Middleditch, J.,
%Kristian, J.A. \& Pennypacker, C.R., ApJ {\bf 476}, 281 (1997)
%\reference{} 14. STScI AGN Atlas (Calibration data http://www.stsci.edu)
\end{references}
\end{document}